\documentclass[graybox]{svmult}

\usepackage{url}
\usepackage{hyperref}
\usepackage{mathptmx}       
\usepackage{helvet}         
\usepackage{courier}        
\usepackage{type1cm}        
\usepackage{wrapfig}                            
\usepackage{amssymb}
\usepackage{mathtools}
\usepackage{makeidx}         
\usepackage{graphicx}        
\usepackage{amsmath}                             
\usepackage{multicol}        
\usepackage[bottom]{footmisc}
\usepackage{romannum}
\usepackage{eucal}
\usepackage{array}
\usepackage{subcaption}
\usepackage{xcolor}
\usepackage{csquotes}
\usepackage{appendix}
\usepackage{float}
\usepackage{graphicx,caption}
\usepackage{geometry}
\makeindex             

\title*{DUNE potential for sub-GeV dark matter in proton beam dump mode}

\author{Sabeeha Naaz, Jyotsna Singh and R. B. Singh}
\institute{Sabeeha Naaz, Dr. Jyotsna Singh and Dr. R. B. Singh \at University of Lucknow, Department of Physics,Lucknow-226007,India 
\email{sabeehanaaz0786@gmail.com}
\email{singh.jyotsnalu@gmail.com}
\email{rajendrasinghrb@gmail.com}}
\begin{document}
\maketitle
\vspace{-25mm}
\abstract:{DUNE with its cutting edge technology is designed to study the neutrino science and proton decay physics. This facility
can be further exploited for the study of the ground breaking discoveries i.e. origin of matter, unification of forces, dark
matter detection etc. In this work we have explored the DUNE potential for capturing the sub-GeV dark matter
in viable dark matter parameter space. The scenario of sub-GeV dark matter range requires a light mediator that couples the hidden sector with the
standard model. The choice of the mediator will decide the different channels by which dark matter candidates can be produced.
Here three channels $\pi^{0}/\eta$-decay, proton bremsstrahlung and parton-level production modes 
are considered for the production of dark matter with a 120 GeV proton beam facility placed at Fermi lab.
To overcome the neutrino background we have used beam dump mode for the production of pure dark matter beam.
To explore the new region of parameter space of dark matter at DUNE the elastic scattering of dark matter beam with electrons and nucleons are 
studied. In terms of DUNE potential for capturing dark matter signatures
the dark matter yield results at DUNE (in our work) shows a significant 
improvement over existing dark matter probes i.e. BaBar, E137, LSND, MiniBooNE, T2K etc.
}


\section{Introduction:}
 The existence of dark matter is strongly suggested by various gravitational phenomena in astrophysics and cosmology. 
 The simplest realization of dark matter particles is in form of new stable weakly interacting non-baryonic elementary particles,
 which can be relativistic or non-relativistic. 
 The presence of such particles has motivated experimentalist to design experiments capable of capturing dark matter (DM) signals.
 The three main experimental techniques which are used to probe the dark matter particles are (\romannum{1}) Indirect Detection  
 experiments \cite{72,73,74} (\romannum{2})
 Direct Detection experiments \cite{71} and (\romannum{3}) Collider searches
 \cite{75,76,77,78}. In direct detection mode
 the dark matter particles collide with the standard model particles and the study of recoil nucleons or scattered electrons provide 
 the information regarding dark matter candidates.
 While the observation of standard model (SM) particles created during the annihilation or decay of dark matter particles will come under indirect 
 detection searches. 
 Further in collider searches dark matter beam is produced by the collision of highly energetic particles.
 
 The mass of  different DM candidates, which are postulated using different inputs of physics varies from 
 $10^{-31}$ to $10^{20}$ GeV. The present research on DM (based on their masses) can be divided
 on the basis of detection technique, type of its interaction and DM mass i.e. axion DM produced from the non-thermal
 mechanism $\mathcal{O}(eV)$, light or sub-GeV DM $\mathcal{O}(MeV - GeV)$ and weakly interacting massive particles (WIMPs) ($ > 1 GeV$) are
 produced from the thermal freeze-out mechanism. Due to absence of the signatures of DM candidates the cosmologist have speculated different 
 types of the DM,  
 whose classification depends on the kinetic energy possessed by the DM.
If the DM particles are relativistic or super relativistic they are called hot DM \cite{105} 
otherwise they are called cold DM. Since the hot dark matter particle are moving with relativistic or super relativistic speed hence they will escape
from small density fluctuations. If the DM particles are moving very fast then 
the structure formation
theory predicts that all structures smaller than massive galaxies would be destroyed by free streaming process. But we have structures smaller than
galaxies so we can assume that DM can not be predominantly hot. In this way we can assume that a little bit of hot DM is mixed with predominantly cold
DM.
 
 In one of the strongly motivated idea, DM particles are considered as the natural partners of the existing standard model particles and originate from
 the unification 
 frameworks or axions invoked to solve the strong CP 
 violation problems. The line of thought followed for the production of DM particles will determine the method or the experimental
 techniques, required for the detection of DM. The sensitivity of these experiments must allow the experimental facility to explore new interaction 
 channels, whose cross-sections are below the cross-section of electroweak interaction of the SM particles.
 With these types of experiments we can try to achieve
 the required relic DM density without depending on either weak scale mass or weak scale interaction.
 The null results estimated by the direct detection technique \cite{79,80,81,82,83} used for the capturing of DM signatures of mass range 
 1 GeV to 1 TeV has ruled out the large region of DM parameter space.
 These results motivate us to move towards the the discovery of production modes, which can produce lower mass DM candidates (below 1 GeV).
 This low mass DM can be produced using various techniques \cite{84}. In our paper, we have considered direct detection 
 technique to probe the sub-GeV DM and the constraints imposed on the mass of DM 
 lies well below the Lee-Weinberg bound \cite{4} which promotes a non SM annihilation channel through light states to produce correct relic 
 abundance \cite{103}.
 
 The characteristics of weak-scale (low mass mediator) parameter offers a hope for direct detection of non-gravitational DM 
 interactions in laboratory.
 However these searches are less sensitive to light DM candidates due to the low detection threshold of recoil nucleon, thus it is 
 essential to consider alternative approach to detect light DM in this regime (sub-GeV range).
 New strategy introduced in above context are fixed target experiments as they can provide the valuable sensitivity required for the detection of
 sub-GeV DM in comparison to the collider experiments. But one major issue with
 fixed target experiments is the presence of high neutrino background which make it difficult to discriminate between the signatures of DM particles and 
 neutrino present in the background. The use of experiment in beam dump mode reduces the neutrino background in the target volume.
 High intensity proton beam or electron beam 
 produces DM particles after colliding with the nuclei of fixed target, further neutrino background is reduced by using it in beam dump mode.
 The signatures of DM are identified by the different scattering modes of DM i.e. its scattering with the
 nucleons and electrons presence in the downstream near detector.
 This beam dump facility was used by MiniBooNE experiment in 2014 \cite{7} to the detection of DM.
 
 In this paper, we have investigated the sensitivity of DUNE (Deep Underground Neutrino Experiment) \cite{1} for light dark 
 matter using a simulation tool BdNMC \cite{61}. The physics used here for DM candidates, considers the interaction of hidden sector 
 (Hypercharge field strength $F_{D\mu\nu}^{Y}$) with standard model (SM) particles (Hypercharge field strength $F_{\mu\nu}^{Y}$) via vector portal.
 A highly energetic proton beam of 120 GeV is bombarded on the fixed target to produce DM beam.
 Different production channels: mesons decay, bremsstrahlung and parton-level are used to produce the DM beam.
 In this work, we have studied the yield of the scattered dark matter at DUNE near detector through the elastic neutral current (NCE) scattering with the
 nucleon and electron to check the sensitivity of DUNE for DM detection.
                
 \vspace{-3mm}
\section{Thermal Relic Light Dark Matter and Dark Matter Portals:}
Thermal relic dark matter are the particles which were created thermally in the early universe, when the temperature 
of the universe was sufficiently high and
thermal equilibrium for DM particle was achieved (i.e. annihilation and creation rates of DM was roughly equal), making dark matter density constant.
As the universe was
cooling and expanding the DM density began to fall down because the DM creation stopped due to cooling but annihilation of the DM continued,
till the time the temperature of the universe was higher than the
mass of DM. When the temperature dropped
below the mass of the DM, the annihilation process of DM also stopped making the DM density 
constant (freeze-out) at this temperature.
 
The cosmological abundance of thermal relic DM can provide the measure of annihilation rate of DM $\langle\sigma v\rangle$ \cite{86}.
The annihilation rate of DM can be expressed as,
\begin{equation}
 \langle\sigma v\rangle = x + yv^{2} +.............
\end{equation}

First term of above equation indicates the annihilation cross-section arising from the s-wave which is independent
of the relative velocity of DM ($v$) whereas the second term 
gives the contribution arising from partial p-wave which depends on the square of relative velocity of DM.
The s-wave contribution dominates in the thermally averaged annihilation cross-section 
($\langle\sigma v\rangle_{s} \approx 2.2 \times 10^{-26}cm^{3}s^{-1}$) but adding p-wave
contribution to it gives precise value of relic DM density. In our model we have considered s-wave and p-wave contributions 
since the detection of annihilation cross-section arising from partial waves other than s-wave will provide a crucial clue 
regarding the nature of DM. However these detection become inefficient for heavy DM (WIMP) candidates.

The lack of WIMP signatures at LHC (Large Hadron Collider) and other experiments diverted the attention from massive DM candidates to lower masses of 
DM candidates. This selection was 
inspired by the observation of 511 keV $\gamma$ -ray line arising from the galactic center with the help of SPI (SPectrometer INTEGRAL) spectrometer in the 
INTEGRAL (INTErnational Gamma-Ray Astrophysics Laboratory) space observatory \cite{43,44}.
It was also inspired by many theoretical models i.e. supersymmetric (susy) model of DM, 
in this supersymmetric model susy partner particle for each SM particle is included and by adding these particles 
the supersymmetric models attempt to address problems,
which are beyond the purview of SM.
The susy model for DM predicts the precise annihilation cross-section of DM which can generate the correct thermal relic DM abundance \cite{50,51,52}.
One more relevant theoretical approach to detect sub-GeV DM is asymmetry DM (ADM) model.  
This ADM model relates the asymmetry of dark matter to the asymmetry of baryon. In the present universe, DM density is approximately 
five times of the SM baryon density \cite{38}. If the production channel for the creation of the dark matter density and baryonic density in the
early universe is assumed to be the same then DM relic density can be naturally explained with the help of ADM model framework \cite{17,18,36,37}.

In an attempt to understand the universe, theoretical physicists have used different portals which establish a connection between the hidden sector 
and visible sector. Portals acts like a bridge between dark matter and ordinary matter and they exchange specific mediator between them. 
The hidden sector mediators can be scalar, pseudoscalar, vector or fermionic. 
The lagrangian of different dominant portals with associated mediators are as follows \cite{84}:

\begin{equation}
\mathcal{L}_{portal} =  \begin{cases}
                         - \epsilon F_{\mu\nu} F_{D}^{\mu\nu} &\text{Vector Portal :: Vector}\\
                         (\mu S + \lambda S^{2}) HH^{\dagger} &\text{Higgs Portal :: Scalar}\\
                         y_{ij} L_{i}HN_{j} &\text{Neutrino Portal :: Fermion}\\
                         \frac{a}{f_{a}}A_{\mu\nu}\tilde{A}^{\mu\nu} &\text{Axion Portal :: Pseudoscalar}\\
                        \end{cases} 
                        \end{equation}

Here, $\epsilon$ is kinetic mixing term, $F_{\mu\nu} = [\partial_{\mu}F_{\nu} - \partial_{\nu}F_{\mu}]$ is the SM hypercharge field strength tensor,
$F_{D_{\mu\nu}} = [\partial_{\mu}\gamma_{D_{\nu}} - \partial_{\nu}\gamma_{D_{\mu}}$] is field strength of dark sector $U(1)_{D}$, $H$ is the SM Higgs doublet, 
$S$ is scalar singlet of new dark sector $U(1)_{D}$, $\lambda$ and $\mu$ are free parameters, $L$ is a lepton doublet, $i$ and $j$ are flavor indices,
$N_{j}$ 
is right handed state for majorana mass term, $f_{a}$ is mass scale of pseudoscalar $a$ and $A_{\mu\nu}(\tilde{A}^{\mu\nu})$ is the
(dual) field strength tensor of the SM photon field.

We have focused on the vector portal which is the most viable portal for the production of light
dark matter because the \enquote{kinetic mixing} interaction term \enquote{$\epsilon F_{\mu\nu} F_{D}^{\mu\nu}$} is invariant under gauge transformation
of both $U(1)_{D}$ and $U(1)_{Y}$ in this scenario.
To consider the DM candidates the SM can be extended by adding
a new extra dark gauge group $U(1)_{D}$ that couples to the SM hypercharge $U(1)$ gauge group via kinetic mixing term $\epsilon$
 \cite{15,16}. 
The dark gauge group $U(1)_{D}$ symmetry is spontaneously broken at a low energy scale by the higgs in hidden sector and leads a massive
dark photon (DP) mediator $\gamma_{D}$ which is a hypothetical particle of $U(1)_{D}$ gauge group. 
Experimental searches of dark photon is needs of great interest for DM physics community and these searches are carried out at LHC and other experiments.
Here we have focused on the MeV to GeV mass range of dark photon because this mass scale of 
dark photon will be able to describe the shocking results of astrophysical evidences regarding DM density.
The DM particle $\chi$ can be taken as scalar or fermionic fields which are charged
under dark gauge group $U(1)_{D}$. We have focused on scalar field of DM candidates 
because this scenario is less constrained by astrophysical considerations, such as annihilation-induced distortion of the CMB.

The applicable low energy Lagrangian of light scalar dark matter for vector portal can be expressed as,
\begin{equation}
 \mathcal{L}_{DM} = \mathcal{L}_{\gamma_{D}} + \mathcal{L}_{\chi}
\end{equation}
where,
\begin{equation}
 \mathcal{L}_{\gamma_{D}} = -\frac{1}{4}F_{D_{\mu\nu}}F_{D}^{\mu\nu} + \frac{1}{2}m_{\gamma_{D}}^{2}\gamma_{D_{\mu}}\gamma_{D}^{\mu}
 - \frac{1}{2}\epsilon F_{D_{\mu\nu}}F^{\mu\nu}
\end{equation}
\begin{equation}
 \mathcal{L}_{\chi} = \frac{i g_{D}}{2}\gamma_{D}^{\mu}J^{\chi}_{\mu} + \frac{1}{2} \partial_{\mu} \chi^{\dagger}\partial_{\mu} \chi - m_{\chi}^{2}\chi^{\dagger}\chi
\end{equation}

where,
$J_{\mu}^{\chi} = [(\partial_{\mu}\chi^{\dagger})\chi - \chi^{\dagger}(\partial_{\mu}\chi)]$ 
is DM current density and $g_{D}$ is treated as gauge coupling constant between DM current density and a new massive dark
photon field potential $\gamma_{D}^{\mu}$ originated from spontaneously broken gauge symmetry $U(1)_{D}$. 
At low energies, DP kinetically mixes with ordinary photon and couples to the SM electromagnetic current with strength
$\epsilon g$, where $g$ is electromagnetic coupling constant.

This minimal light dark matter model consist of four parameters, dark matter mass ($m_{\chi}$), 
dark photon mass ($m_{\gamma_{D}}$), gauge coupling ($g_{D}$) and kinetic mixing term ($\epsilon$). 
The viable sub-GeV scale DM under the Lee-Weinberg bound provides a fix relation among all parameters 
and can give correct relic DM density \cite{11}.

Our interested region of light DM is that in which DM is thermal relic. 
Sub-GeV thermal relic DM candidates are constrained by annihilation cross-section $\langle\sigma v\rangle \sim 1 pb$ in which thermal relic DM
abundance are compatible with the observed relic DM density. Many experimental constraints are present for the selection of 
parameter
space of light DM i.e. strongest constraints on DM-electron scattering are imposed by LSND experiment \cite{59,c,88,95},
CRESST-II \cite{93} direct detection
search places constraints on the large value of fine structure parameter ($\alpha_{D}$) of hidden sector, BaBar \cite{k} experiment is mono-photon 
searches provides constraint on the mass of DM is $m_{\chi} > 60 MeV$, beam dump experiment E137 \cite{96} is sensitive for light DM with 
20 GeV electron beam and NA64 collaboration \cite{l} recently provides a strong constraints on the mass of dark photon which should be 
below the 100 MeV.

The reachable parameter space by neutrino facilities is $m_{\gamma_{D}} > 2m_{\chi}$ and $g_{D} \gg \epsilon g$ 
which suggest that dark photon mostly decays into dark matter pair and for this situation the annihilation cross-section of scalar DM into 
leptons is discussed in reference \cite{64}:
\begin{equation}
 \sigma (\chi\chi^{\dagger} \rightarrow l \overline{l})v \sim \frac{8\pi v^{2}Y}{m_{\chi}^{2}}
 \label{6}
\end{equation}
where,
\begin{equation}
 Y = \epsilon^{2}\alpha_{D}\left(\frac{m_{\chi}}{m_{\gamma_{D}}}\right)^{4};     \alpha_{D} = \frac{g_{D}^{2}}{4\pi}
 \label{7}
\end{equation}
and $v$ is relative velocity of DM. The DM annihilation cross-section depends of $Y$ which in turn depends on the
$\epsilon$, $\alpha_{D}$, $m_{\chi}$ and $m_{\gamma_{D}}$ parameters then
we have tried to explore the correlation between $(Y, m_{\chi})$ and $(\epsilon, m_{\gamma_{D}})$ for DUNE detector.
\section{Production of the Dark Matter Candidates:}
Several models have been developed to study the light DM candidates. Amongst these models we have to select one, which contains in itself a viable
annihilation channels to be studied at the selected experiment.
We have selected a dark matter physics model which works for light DM candidates and can be used for DM beam produced at fixed target neutrino experiments.
Near detector of a fixed target neutrino experiments are designed to detect the neutrino signatures but here we want to check their 
sensitivity for DM detection \cite{88}.
The idea behind the DM detection by the fixed target neutrino experiments is as follows: a highly intense proton beam interacts with the fixed target and 
produces dark photons which further decays into
DM candidates. In this way beside neutrino beam, a DM beam is also produced in neutrino experiments. 
In the presence of neutrino beam the detection of DM particles will be difficult as the neutrino beam will act as a background and this background will
add huge errors in the detection of DM. Therefore beam dump mode is selected for exploring the sensitivity of DUNE. In this mode proton beam interacts
with the beryllium target and produces charged pions which are 
aligned using magnetic horns towards a steel absorber (beam dump) where these pions decay into neutrinos. In this way nearly pure
DM beam is produced in beam dump mode.

Here most of the light dark photons are produced from the decay of neutral 
mesons. Some heavy dark photons are also produced from the resonance vector mesons and from the hadron-level interactions. 
Keeping in mind that beam energy of DUNE flux is high here we summarize the three production modes of light dark matter.
\begin{enumerate}
 \item $\pi^{0}/\eta$  mesons decay: relevant for lower masses of dark photon.
 \item Resonant vector mesons (Bremsstrahlung) decay: relevant for intermediate masses of dark photon.
 \item Direct or parton-level production from quarks and gluon constituents: relevant for higher masses
 of dark photon ($m_{\gamma_{D}}\textgreater1$ GeV).
\end{enumerate}

\subsection{ $\pi^{0}/\eta$  mesons decay in flight:}
At lower energies the neutral mesons decay channel dominates over other channels for the over all production of dark photon. 
These mesons are produced from
the primary interaction of $p(p)$ and $p(n)$. 
\begin{equation}
 p + p(n) \to X + \pi^{0},\eta \to X + \gamma + \gamma_{D} \to X + \gamma + \chi + \chi^{\dagger}
\end{equation}
This DM production
process depends on the beam energy and configuration of target because decay of mesons can happen either inside the target or in successive 
decay volume. 
Since the mesons distribution varies with the beam energy hence different experiment needs different analytical fit $f(\theta, p)$ for mesons distribution
in $\theta$ and $p$ plane, to generate the distribution of DM flux.
For proton beam energy considered in this work (120 GeV for DUNE) the analytical distribution for mesons considered is BMPT (Beryllium Material 
Proton Target) distribution \cite{54}. Using this distribution in appropriate energy range of mesons we get 
$\sigma_{pp \to pp\pi^{0}} \approx 27\sigma_{pp \to pp\eta}$ \cite{a}.

If the mass of neutral meson is greater than the mass of the dark photon i.e. $m_{\gamma_{D}} < m_{\delta}$ $(\delta = \pi^{0},\eta)$ or if the mass of
DM particle produced is such that $2m_{\chi} < m_{\gamma_{D}} < m_{\delta}$ then $\gamma_{D}$ will be produced on-shell and it will decay into DM candidates. 
Branching ratio of mesons decay to dark matter using narrow width approximation \cite{67} is equal to the product of 
mesons decay to dark photons and dark photon to
dark matter. The production of dark photon depends on $\epsilon^{2}$ and ratio of $m_{\gamma_{D}}$ and $m_{\delta}$ while it is
independent of $m_{\chi}$ and $\alpha_{D}$.

\begin{equation}
Br(\delta \to \gamma\chi\chi^{\dagger}) = Br(\delta  \to \gamma\gamma_{D}) Br(\gamma_{D} \to \chi\chi^{\dagger}) \\
\end{equation}
\begin{equation}
Br(\delta \to \gamma\gamma_{D}) \simeq 2\epsilon^{2} \left(1 - \frac{m_{\gamma_{D}}^{2}}{m_{\delta}^{2}}\right)^{3} 
 Br(\delta \to \gamma \gamma)
\end{equation}

Where $Br(\pi^{0} \to \gamma \gamma) \simeq 1 $ and $Br(\eta \to \gamma \gamma) \simeq 0.39 $.
For completeness, the dark photon always decay into DM indicates; $Br(\gamma_{D} \to \chi \chi^{\dagger}) \simeq 1$,
which requires $\epsilon \ll 1$ condition.

For off-shell production of dark photon i.e. $2m_{\chi} > m_{\gamma_{D}}$ or $m^{2}_{\gamma_{D}} \gtrsim
m_{\delta}^{2} - 2\Gamma_{\gamma_{D}}m_{\gamma_{D}}$ (where $\Gamma_{\gamma_{D}}$ is decay width of DP), the narrow width approximation is not a viable
choice. 
This production mode does not have any analytical form like in the case of on-shell production. Here the DM production is calculated by 
by three body decay
and branching ratio,
branching ratio for off-shell dark matter production is expressed as \cite{68},
\begin{equation}
 Br(\delta \to \gamma\chi\chi^{\dagger}) = \frac{\epsilon^{2}\alpha_{D}}{\Gamma_{\delta}} \times \frac{1}
 {4\pi m_{\delta}} \int d\Psi_{\delta \to \gamma\gamma_{D}} d\Psi_{\gamma_{D} \to \chi\chi^{\dagger}} dq_{\gamma_{D}}^{2} 
 |A_{\delta \to \gamma\chi\chi^{\dagger}}|^{2}
\end{equation}
where $d\Psi$ is two body phase-space, $q_{\gamma_{D}}^{2}$ is momentum of DP, 
$\Gamma_{\delta}$ is total decay width of $\pi^{0}$ and $\eta$, and $A_{\delta \to \gamma\chi\chi^{\dagger}}$ is three body decay 
normalized amplitude.\\
\begin{equation}
|A_{\delta \to \gamma\chi\chi^{\dagger}}|^{2} = \frac{\epsilon^{2}\alpha_{D}\alpha^{2}}{\pi f_{\delta}^{2}[(q_{\gamma_{D}}^{2} - 
m_{\gamma_{D}}^{2})^{2}
  + m_{\gamma_{D}}^{2}\Gamma_{\gamma_{D}}^{2}]} [(q_{\gamma_{D}}^{2} -4 m_{\chi}^{2})(m_{\delta}^{2}
  - q_{\gamma_{D}}^{2})^{2} - 4q_{\gamma_{D}}^{2}(p.k_{1} - p.k_{2})^{2}]
\end{equation}
Where, $f_{\delta}$ is meson ($\pi^{0}, \eta$) decay constant, $p$ and $q_{\gamma_{D}}$ are momenta of ordinary photon and dark photon
respectively, $k_{1}$ and $k_{2}$ are momenta of dark matter and anti dark matter produced in the final state.

The total dark matter particles produced via both on-shell and off-shell production can be written as,\\
\begin{equation}
 N_{\chi} = \sum_{on-shell + off-shell} Br(\delta \to \gamma\chi\chi^{\dagger}) \times \delta\textunderscore per\textunderscore\pi^{0} 
 \times \pi^{0}\textunderscore per\textunderscore POT \times POT
\end{equation}
where, $\delta\textunderscore per\textunderscore\pi^{0} = 1$  for $\pi^{0}$ and $\delta\textunderscore per\textunderscore\pi^{0} = 0.11$ for $\eta$ 
which gives the $\pi^{0}$ production 27 times that of $\eta$ production.
$\pi^{0}\textunderscore per\textunderscore POT$ and which is number of $\pi^{0}$ per proton on target (POT) is equal to $4.5$ \cite{b} for DUNE.

\subsection{Resonant vector mesons (proton Bremsstrahlung) decay:}
Proton bremsstrahlung production process becomes effective for intermediate mass range of mediators. The dark photon produced in this process is via 
the scattering of proton by nucleons of fixed target which is similar to the ordinary proton bremsstrahlung. \\
\begin{center}
$p + p(n) \to p + p(n) + \gamma_{D}, $
\end{center}
In this process a nearly collimated beam of dark photon is generated. The four momentum assigned to incident proton of mass $m_{p}$ is 
$q = (E_{p}, 0, 0, Q)$ where $E_{p} = Q + \frac{m_{p}^{2}}{2Q}$. The four momentum of outgoing dark photon of mass $m_{\gamma_{D}}$ is 
$q_{\gamma_{D}} = (E_{\gamma_{D}}, q_{\perp}\cos(\phi),
q_{\perp}\sin(\phi), Q.z)$ 
where $E_{\gamma_{D}} = Q.z + \frac{q_{\perp}^{2} + m_{\gamma_{D}}^{2}}{2Q.z}$, $Q.z = q_{\parallel}$ and z is a fraction of proton beam momentum 
carried away by outgoing DP in the direction of proton beam. Here $q_{\perp}$ and $q_{\parallel}$ are transverse and longitudinal components 
of $\gamma_{D}$ momenta.\\
By Weizs$\ddot{a}$cker-Williams
approximation the rate of dark photon production per proton is as follows \cite{56,57},\\
\begin{equation}
 \frac{d^{2}N_{\gamma_{D}}}{dz dq_{\perp}^{2}} = 
 \frac{\sigma_{pA}(2m_{p}(E_{p} - E_{\gamma_{D}}))}{\sigma_{pA}(2m_{p}E_{p})}F^{2}_{1, N}(q^{2})f_{yx}(z, q_{\perp}^{2})
 \label{14}
\end{equation}
\\
here $\sigma_{pA} = f(A)\sigma_{pp}$, $f(A)$ is a function of atomic number $A$ and $f_{yx}(z, q_{\perp}^{2})$
is a splitting weight-function of photon which relates before and after scattering differential cross section \cite{56},\\

$f_{yx}(z, q_{\perp}^{2}) = \frac{\epsilon^{2}\alpha}{2\pi H}\left[\frac{1 + (1 - z)^{2}}{z} - 2z(1-z)\left
(\frac{2m_{p}^{2} + m_{\gamma_{D}}^{2}}{H} - z^{2}\frac{2m_{p}^{4}}{H^{2}}\right)
+ 2z(1 - z)(z + (1 - z)^{2})
\frac{m_{p}^{2}m_{\gamma_{D}}^{2}}{H^{2}} + 2z(1 - z)^{2}\frac{m_{\gamma_{D}}^{4}}{H^{2}} \right]$ 
\\
here $H = q_{\perp}^{2} + (1 - z)m_{\gamma_{D}}^{2} + z^{2}m_{p}^{2}$ and $\alpha$ is electromagnetic fine structure constant.

Since radiative $\gamma_{D}$ has time-like momentum and
time-like form factor $F_{1,N}(q^{2})$ expresses off-shell mixing with vector mesons in appropriate kinematic region.
Vector portal consider only proton form factor $F_{1,p}(q^{2})$ \cite{58} which incorporates both $\rho$-like (isovector) and $\omega$-like 
(isoscalar) Breit-Wigner components \cite{58} and this form factor is not completely resolved for $\rho$ and $\omega$. 
Above 1 GeV, the form factor suppresses the rate of production of virtual photons and at this moment direct parton level
production comes into play.

To calculate the dark photon production rate, equation [\ref{14}] must be integrated over $p_{\perp}$ and $z$ for a range that satisfies
some kinematic conditions \cite{56}:
\begin{equation}
 E_{p}, E_{\gamma_{D}}, E_{P} - E_{\gamma_{D}} \gg m_{p}, m_{\gamma_{D}}, \left|{q_{\perp}}\right|
\end{equation}
We have selected a range $z \in [0.2,0.8]$ and $\left|{p_{\perp}}\right| = 0.4$ for DUNE which satisfied the above conditions.
                 
\begin{figure}[H]
\centering
\begin{subfigure}{.4\textwidth}
  \centering
  \includegraphics[width=1.0\linewidth]{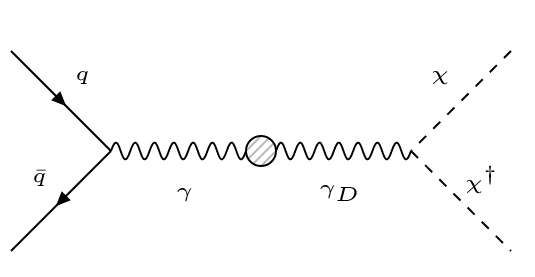}
  \end{subfigure}%
\begin{subfigure}{.4\textwidth}
  \centering
  \includegraphics[width=1.0\linewidth]{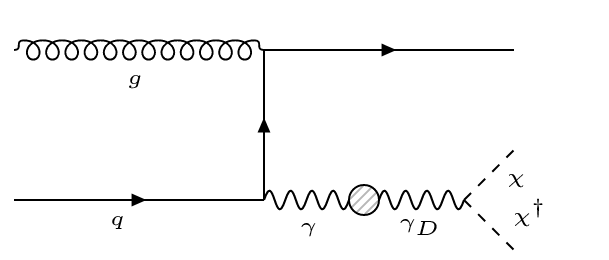}
 \end{subfigure}
\caption{Parton-level production of scalar dark matter via vector portal.}
\label{fig1}
\end{figure}
                 
\subsection{Direct production of dark matter from Drell-Yan process:}
Parton-level production mechanism is very interesting as it provides a portal for the production of high energetic dark photon.  
Direct production of dark photons from Drell-Yan process is shown in figure [\ref{fig1}].
\begin{equation}
 p + p(n) \to X + \gamma_{D} \to X + \chi\chi^{\dagger}
\end{equation}
This process will use the narrow width approximation for on-shell DM production \cite{59} and this approximation is valid up to the second order.
The dark matter production cross-section as a function of lab frame
DM energy $E_{\chi}$ and angle between the direction of proton beam and lab frame momentum of DM $\phi$ can be expressed as,
\begin{equation}
 \frac{d^{2}\sigma(p + p(n) \to X + \gamma_{D} \to X + \chi\chi^{\dagger})}{dE_{\chi}d\cos{\phi}} = 
 \left[\frac{\partial(x,\cos{\hat{\phi}})}{\partial(E_{\chi},\cos{\phi})}\right] \times \frac{d\sigma(pp(n) \to \gamma_{D})}{dz}
 Br(\gamma_{D} \to \chi\chi^{\dagger})g(\cos{\hat{\phi}})
 \label{17}
\end{equation}
where $\hat{\phi}$ is angle between momentum of dark matter candidate and proton beam, in the dark photon rest frame, and square bracket term is a
Jacobian function. The $g$ function gives the angular distribution of the dark matter in $\gamma_{D}$ rest frame for scalar DM produced via a vector
mediator and it can be expressed as,
\begin{equation}
 g(\cos\hat{\phi}) = \frac{3}{4}(1-\cos^{2}\hat{\phi})
\end{equation}
the direct production cross-section of dark photon is \cite{59},\\
\begin{equation}
 \sigma(pp(n) \to \gamma_{D}) = \int_{\zeta}^{1} \frac{d\sigma(pp(n) \to \gamma_{D})}{dx} dx = 
 \frac{4\pi^{2}\alpha\epsilon^{2}}{m_{\gamma_{D}}^{2}}
 \sum_{q}^{}Q_{q}^{2} \int_{\zeta}^{1} \frac{dx}{x} \tau \left[f_{q/p}(x)f_{\overline{q}/p(n)}(\frac{\zeta}{x}) 
 + f_{\overline{q}/p}(x)f_{q/p(n)}(\frac{\zeta}{x})\right]
\end{equation}
where $Q_{q}$ is quark charge in the unit of positron electric charge, $\zeta = m_{\gamma_{D}}^{2}/s$ and $\sqrt{s}$ is the hadron-level
center of mass energy. $f_{q/p(n)}(x)$ is the parton distribution function (PDF) which gives the probability of extraction of
quarks and gluons from proton(neutron) with longitudinal momentum fraction $x$. 
To evaluate cross-section, we have used CTEQ6.6 PDFs \cite{60} and have set $Q = m_{\gamma_{D}}$ which is varied in range of
$m_{\gamma_{D}}/2$ to $2m_{\gamma_{D}}$. \\
To produce the dark matter we integrate the equation [\ref{17}] over the DM lab frame energy and angle between proton beam direction and 
momentum of DM lab frame.

\section{Scattering and Cross-Section of Dark Matter Beam:}

In this work we explore the allowed parameter space for capturing dark matter signatures at DUNE experiment through elastic neutral current-like 
(NCE-like) scattering of DM candidate with nucleons and electrons.

\begin{figure}[H]
\centering
\includegraphics[width=0.5\linewidth]{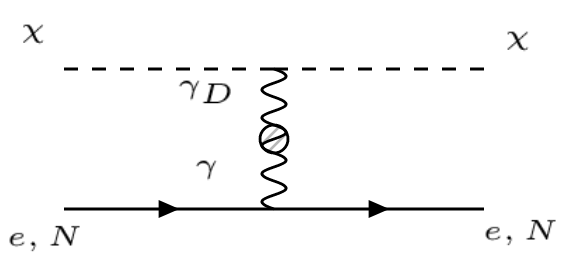}
\caption{Tree-level DM scattering with nucleons and electrons via vector portal.}
\label{fig2}
\end{figure}  

\subsection{Elastic neutral current-like scattering with electrons:}
The differential cross-section of DM elastic scattering with electrons \cite{c} as a function of scattered
electron energy ($E_{e}$) is given below:
\begin{equation}
 \frac{d\sigma_{\chi e\to \chi e}}{dE_{e}} = 4\pi \alpha\alpha_{D}\epsilon^{2} \times \frac{2m_{e}E^{2} - 
 (2m_{e}E + m_{\chi}^{2})(E_{e} - m_{e})}{(E^{2} - m_{\chi}^{2})(m_{\gamma_{D}}^{2} + 2m_{e}E_{e} - 2m_{e}^{2})^{2}}
 \label{21}
\end{equation}
where $m_{e}$ is mass of the electron, $m_{\chi}$ is mass of the dark matter and $E$ is the energy of incoming dark matter.
\subsection{Elastic neutral current-like scattering with nucleons:}
Scattering of scalar DM with nucleons via vector mediator is similar to the neutrino-nucleon scattering via $Z$-mediator \cite{62}.
The differential cross-section of incoherent elastic scattering of dark matter with free nucleons with 
respect to the outgoing dark matter energy ($E_{\chi}$) can be expressed as \cite{c,d},
\begin{equation}
\frac{d\sigma_{\chi N\to \chi N}}{dE_{\chi}} = 4\pi \alpha\alpha_{D}\epsilon^{2} \times 
\frac{F_{1,N}^{2}(Q^{2})A(E,E_{\chi}) - \frac{1}{4}F_{2,N}^{2}(Q^{2})B(E,E_{\chi})}{(m_{\gamma_{D}}^{2} + 2m_{N}(E - E_{\chi}))^{2}
(E^{2} - m_{\chi}^{2})}
\end{equation}

where $E$ and $E_{\chi}$ represents energy of incoming and outgoing dark matter and
$Q^{2} = 2m_{N}(E - E_{\chi})$ is the transferred four momentum. The function A and B are expressed as:
\begin{equation}
 A(E,E_{\chi}) = 2m_{N}EE_{\chi} - m_{\chi}^{2}(E - E_{\chi});      \\
 B(E,E_{\chi}) = (E_{\chi} - E)[(E_{\chi} + E)^{2} + 2m_{N}(E_{\chi} - E) - 4m_{\chi}^{2}]
\end{equation}
The contribution to this cross-section arises from both nucleons: protons and neutrons hence their form factors are present in the above equation
[\ref{21}].
The $F_{1,N}$ and $F_{2,N}$ represents monopole and dipole form factors of nucleons respectively and can be expressed as,  
\begin{equation}
 F_{1,N} = \frac{q_{N}}{(1 + Q^{2}/m_{N}^{2})^{2}};
 F_{2,N} = \frac{\kappa_{N}}{(1 + Q^{2}/m_{N}^{2})^{2}}
\end{equation}
where $q_{p}$ = 1, $q_{n}$ = 0, $\kappa_{p}$ = 1.79, and $\kappa_{n}$ = -1.9.
Above scattering equation stands only for free nucleons but we know that in a nucleus bound nucleons also exist. Therefore we can use
effective differential cross-section with a consideration of bound nucleons. The effective differential cross-section can be written as,
\begin{equation}
 \frac{d\sigma_{\chi N\to \chi N}^{eff}}{dE_{\chi}} = \left[\frac{1}{7}C_{pf}(Q^{2}) + \frac{3}{7}C_{pb}(Q^{2})
 \right]\frac{d\sigma_{\chi p\to \chi p}}{dE_{\chi}} +  \frac{3}{7}C_{nb}(Q^{2})\frac{d\sigma_{\chi n\to \chi n}}{dE_{\chi}}
\end{equation}
where C's are the relative efficiencies of dark matter scattering with free proton ($C_{pf}$), bound proton ($C_{pb}$),
bound neutron ($C_{nb}$) and they are the function of momentum transfer $Q^{2}$.

\section{Simulation Technique:} 
We have used BdNMC (Beam dump Neutrino Monte Carlo) \cite{61} simulation tool to find out the sensitivity of DUNE for DM detection.
The DM beam is generated via three different modes; (\romannum{1}) Mesons decay: essential parameters for meson decay channel considered in our work are
$\delta\textunderscore per\textunderscore\pi^{0} = 1$  for $\pi^{0}$, $\delta\textunderscore per\textunderscore\pi^{0} = 0.11$ for $\eta$ and
$\pi^{0}\textunderscore per\textunderscore POT = 4.5$ \cite{b}. 
(\romannum{2}) Bremsstrahlung decay: in proton bremsstrahlung mode DM beam is produced through the decay of $\rho$ and $\omega$ resonant vector mesons. 
(\romannum{3}) parton-level: for parton-level process we have used CTEQ6.6 parton distribution function to generate DM beam.
The details of DM beam production is mentioned in section(3). The BMPT distribution function is used to produce 
DM beam through above three modes. 
A highly intense proton beam
interacts with beryllium target in the fixed target DUNE experiment and produces DM particles via all relevant channels.
These DM particles propagate through decay volume and reach the DUNE near detector.
A brief description of DUNE parameters considered in our work are listed below \cite{63}:

\begin{table}
\begin{center}
\begin{tabular}{ |c|c|c|c|c|c|c|c| } 
 \hline
 Name & Target material & $E_{beam}$ & POT & Detector mass (Fiducial) & Distance & Angle & Efficiency($\epsilon_{eff}$) \\ 
 \hline
 DUNE & $Ar+CH_{4}$ & 120 GeV & 1.1$\times 10^{21}$ & 1 ton(900 Kg + 100 Kg) & 574 m & 0 & 0.9 \cite{66} \\   
 \hline
\end{tabular}
\caption{Essential parameters for DUNE near detector.} 
\end{center}
\end{table}

Elastic scattering of DM candidates by electron and nucleons are studied in our work.
The expression for the combination of total DM signal events observed is taken as \cite{61},
\begin{equation}
 N_{\chi A\to\chi A} = N_{A} \epsilon_{eff} \sum_{p.c.} 
 \left(\frac{N_{\chi}}{2N_{trials}}\sum_{i}L_{i}\sigma_{\chi A,i}\right)
\end{equation}
where $\epsilon_{eff}$ is the detection efficiency of detector, $N_{A(e,p,n)}$ is the electron and nucleon density of target atoms in 
the detector and p.c. mentions the relevant production channel of DM. The $N_{trials}$ is number of trajectories generated by the
Monte Carlo. The inner summation is over 4-momenta of dark matter produced by the Monte Carlo and $L_{i}$ is the 
length of the intersection between dark matter trajectory (4-momentum) and the detector. The DM total cross-section $\sigma_{\chi A}(E)$ is taken as,
\begin{equation}
 \sigma_{\chi A}(E) = \int_{E_{\chi}^{min}}^{E_{\chi}^{max}} dE_{\chi} \sum_{A=e,p,n} f_{A}
 \frac{d\sigma_{\chi,A}}{dE_{\chi}}
\end{equation}
where $E$ is the energy of incoming dark matter, $E_{\chi}$ is the energy of outgoing dark matter and $E_{\chi}^{min/max}$ 
is minimum and maximum outgoing DM energy which is calculated by the 
experimental cuts on the electron/nucleon
recoil momentum $q$, which can be expressed as $\sqrt{2M(E - E_{\chi})}$. 
For vector portal in elastic or quasi-elastic scattering we take $f_{p,n,e}$ = Z, A - Z, Z.\\

\begin{figure}[H]
\centering
\begin{subfigure}{.5\textwidth}
  \centering
  \includegraphics[width=1.0\linewidth]{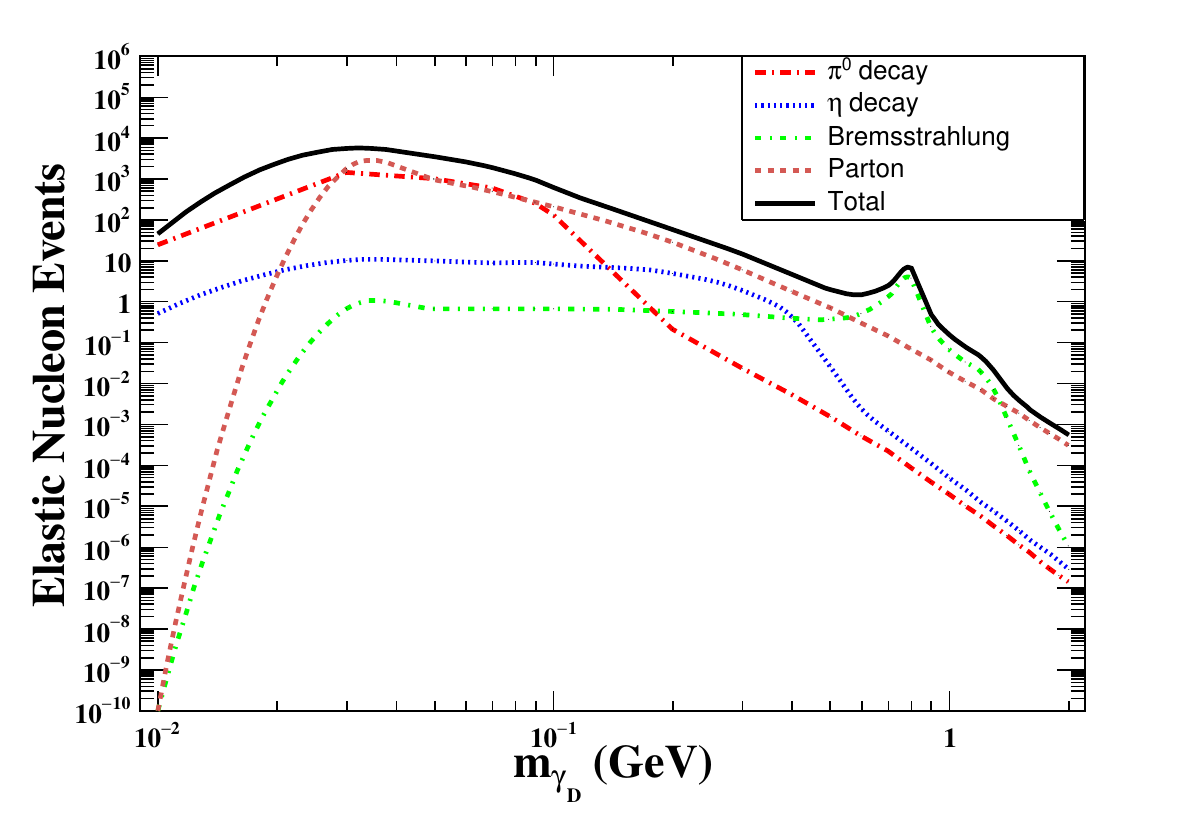}
\end{subfigure}%
\begin{subfigure}{.5\textwidth}
  \centering
  \includegraphics[width=1.0\linewidth]{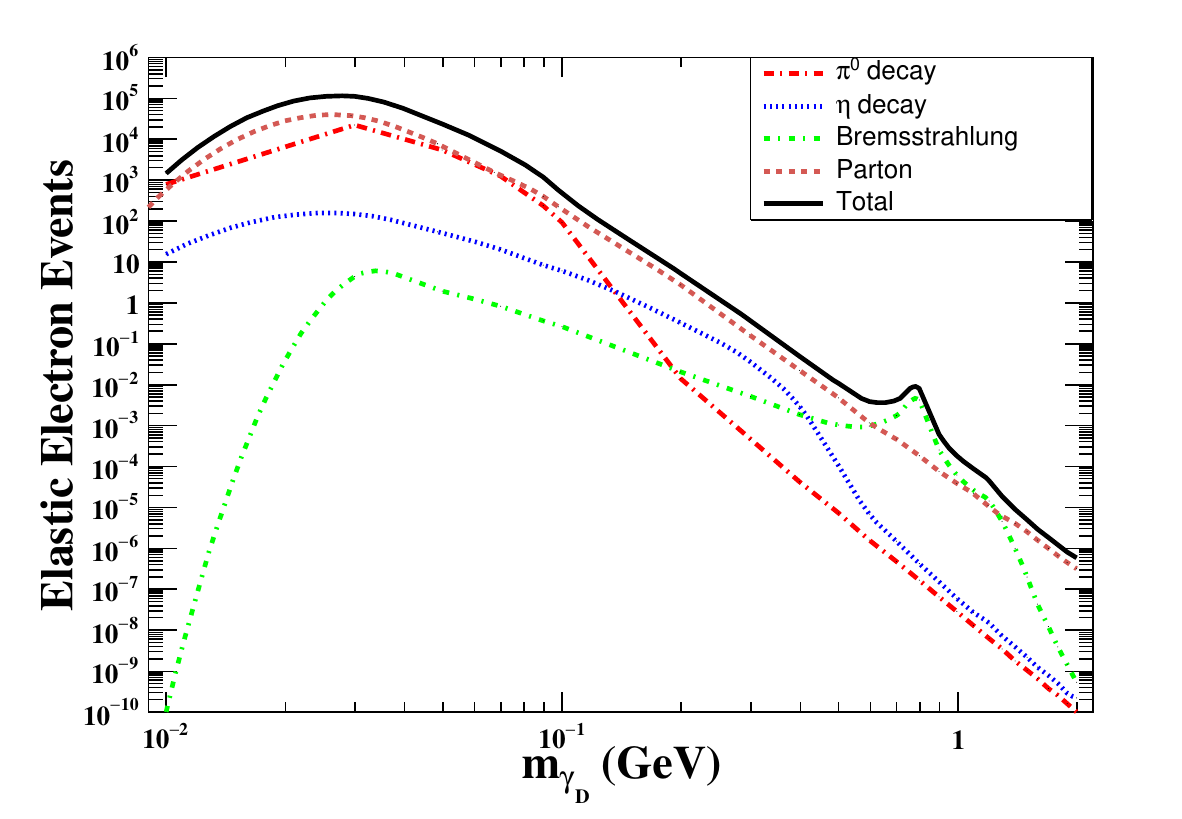}
  \end{subfigure}
\caption{The event rate of DM elastic scattering with nucleons (left panel) and electrons (right panel) versus 
dark photon mass from distinct channels by using 120 GeV beam energy. Here we have considered $m_{\chi}$ = 0.01 GeV, $\epsilon$ = $10^{-3}$ and $\alpha_{D}$ = 0.5.}
\label{fig3}
\end{figure}

\section{Constraints on Minimal Light Dark Matter Model:}
In our work we have considered the coupling of hidden sector with SM particles via vector portal. A benchmark model of light DM is required to 
describe the coupling of hidden sector with SM.
To make the considered model viable with the detection technologies several constraints needs to be imposed on the parameters of the model, which are
summarized below in brief.
\begin{itemize}
 \item \textit{\textbf {Thermal relic DM constraints:}} The constraints imposed by WMAP (Wilkinson Microwave Anisotropy Probe) on the thermal relic DM 
 density, $\Omega_{DM}h^{2} \sim 0.1 \sim (0.1 pb)/\langle \sigma v\rangle_{fo}$ ($h= 0.710\pm0.025$ is Hubble constant in the unit
 of 100 km $sec^{-1} Mpc^{-1}$) introduces the DM annihilation cross-section at 
 freeze-out to be $\langle \sigma v\rangle_{fo} \sim 1pb$. Generally, if DM annihilation cross-section is
 $\langle \sigma v\rangle_{fo} \gtrsim 1pb$ then it does not provide enough contribution to the relic DM abundance.
 \item \textit{\textbf {CMB and BBN constraints on light DM:}} The constraints on DM annihilation cross-section are introduced by the CMB too. 
 The cross-section considered should be such that they do not distort the CMB due to energy injection. The constraint imposed by WMAP7 on the DM 
 annihilation cross-section is $f(z)\langle \sigma v\rangle_{CMB} \lesssim 0.1(m_{\chi}/GeV)pb$,
 here $f(z)$ is an efficiency factor which depends on the redshift $z$. For the lower masses of DM, efficiency factor $f$ varies from $f \sim 0.2$ 
 for pion in the final state to $f \sim 1$ for electron in the final state \cite{87}\cite{59}.
 These constraints excludes the thermal relic DM abundance below the few GeV DM masses via s-channel annihilation. The DM annihilation cross-section via 
 is s-wave (thermally averaged annihilation cross-section is independent of time) suppressed and therefore have less impact on BBN (Big Bang Nucleosynthesis) \cite{97}.
 \item \textit{\textbf {Bullet cluster and cluster lensing constraints:}} The constraint on DM self scattering cross-section imposed by 
 bullet cluster and cluster lensing observations \cite{88a}\cite{89} is given by,
 \begin{equation}
  \frac{\sigma}{m_{\chi}} \lesssim few \times \frac{cm^{2}}{g}
 \end{equation}
 As it is obvious from equation [\ref{6}] and equation [\ref{7}] that above defined constraint will impose  
 bound on the fine structure constant $\alpha_{D}$ of the
 hidden sector at lower DM masses \cite{64}.
 \begin{equation}
  \alpha_{D} \lesssim 0.06 \times \sqrt{\left(\frac{MeV}{m_{\chi}}\right)} \times \left(\frac{m_{\gamma_{D}}}{10 MeV}\right)^{2}
 \end{equation}
  In our work we have used the benchmark value of $\alpha_{D} = 0.5$.
 \item \textit{\textbf {Constraints on the kinetic mixing term $\epsilon$:}} Kinetic mixing parameter $\epsilon$ defines the kinetic mixing of dark photon of 
 hidden sector with ordinary photon of SM via vector portal. Recent bound on $\epsilon$ are set by invisible decay \cite{91, 104} and supernovae \cite{92}.
 The invisible 
 decay of $J/\psi$ (e.g. $J/\psi \to \gamma_{D}^{*} \to \chi\chi^{\dagger}$) and $\Upsilon(1S)$ decay 
 (e.g. $\Upsilon(1S) \to \gamma_{D}^{*} \to \chi\chi^{\dagger}$) imposes constraints on the upper limit, $\alpha_{D}\epsilon^{2} \lesssim few \times 10^{-4}$ for 
 lower DM masses. While the lower limit is imposed by the observation of supernovae which is $\alpha_{D}\epsilon^{2} \gtrsim few \times 10^{-14}$.
 \item \textit{\textbf {Direct detection constraints:}} The recent constraints to the direct detection of light DM are imposed by CRESST-II \cite{93} and 
 CDMS-Lite \cite{94}. The CRESST-II experiment can explore the sensitivity of DM masses below 0.5 GeV with detection threshold of nuclear recoil 307 eV. 
 Whereas CDMS-lite experiment can detect the electron recoils as low as 56 eV for DM masses $\mathcal{O}$(MeV-GeV) hence DM-electron scattering can reach 
 towards lower masses of DM. 
 \item \textit{\textbf {Beam dump experiment constraints:}} LSND fixed target proton beam dump experiment provides a way to investigate sub-GeV DM. 
 This experiment suggests the strongest bounds for on-shell DM production mode $m_{\gamma_{D}} \lesssim m_{\pi^{0}}$ via 
 $p + p(n) \to X + \pi^{0} \to X + \gamma + \gamma_{D} \to X + \gamma + \chi + \chi^{\dagger}$ \cite{59,c,88,95}, these DM candidates when 
 scattered through electrons and nucleons 
 looks similar to neutrino neutral-current scattering signature. The E137 fixed target electron beam dump experiment \cite{96} is sensitive for
 $e + p(n) \to e + p(n) + \gamma_{D} \to e + p(n) + \chi + \chi^{\dagger}$ in a downstream detector. The 
 DM production yield depends on $Y$, \enquote{$Y$} is expressed as $Y = \epsilon^{2}\alpha_{D}\left(\frac{m_{\chi}}{m_{\gamma_{D}}}\right)^{4}$,
 scales with $\epsilon^{2}\alpha_{D}$ and 
 the ratio of $m_{\chi}$ and $m_{\gamma_{D}}$.
 The ratio, $m_{\chi}/m_{\gamma_{D}} = 1/3$ is most conservative value for the estimation of the DM yield, the larger value of ratio will call stronger
 constraints for the estimation of the DM yield.
 \item \textit{\textbf{BaBar experiment:}} A mono-photon search performed at
 BaBar experiment \cite{k} imposes constraint on the mass of dark photon which further invisibly decays
 into DM.
 This experiment eliminates the existence of dark photon with the limits $m_{\gamma_{D}} \textless 8$ GeV and $\epsilon > 10^{-3}$ for $m_{\chi} > 60$ MeV.
 \item \textit{\textbf{$K^{+} \to \pi^{+}\nu\bar{\nu}$:}} The result of E949 experiment at Brookhaven National Laboratory places a limit on kinetic
 mixing parameter $\epsilon^{2} > 3 \times 10^{-5}$ for MeV-GeV mass scale of dark photon \cite{m, n}.
 \item \textit{\textbf{Electron/Muon $g-2$:}} The blue band shown in the figure [\ref{ncesens}] [\ref{esens}] [\ref{nceyield}] and [\ref{eyield}] 
 is the region where the consistency of
 theoretical and experimental value improves to with in 3$\sigma$. Leaving this space all other parameter space is excluded as it increases the 
 disagreement of either muon or electron $g - 2$ to more than 5$\sigma$ \cite{o, p, q}.
\end{itemize}

\section{Results and Discussion:}
World wide there are various proposed and ongoing neutrino experiments whose potential needs to be checked for the existence of light DM.
The sensitivity for the detection of the signatures of DM candidate depends on beam energy; size, geometry and material of 
the detector; angle acceptance of detector
and mitigation of neutrino background. In this work we have explored the sensitivity of DUNE near detector
for the detection of sub-GeV DM with a physics model which is based on the production of DM at a given experiment via vector portal. 
In this model the dark photon of 
gauge group $U(1)_{D}$ kinetically mixes with photon of $U(1)$ gauge group of visible sector. Here the $Y$ parameter is defined in equation [\ref{7}] and it
depends on the kinetic 
mixing term $\epsilon$, fine structure constant of hidden sector $\alpha_{D}$, mass of the DM $m_{\chi}$ and mass of the dark photon $m_{\gamma_{D}}$.
Initially the rate of the rate of DM scattering events with electron and nucleon are checked for different DM production processes taken into 
consideration. These scattering events are plotted as a function of DP mass $m_{\gamma_{D}}$ are shown in figure [\ref{fig3}]. The DP mass is 
allowed to vary from $0.03$ to $2$ GeV.
The DM candidate produced via $\pi^{0}$ channel gives 
a dominant contribution to the scattering events at lower masses of dark photon. Whereas the scattering events of DM candidates produced via bremsstrahlung
and parton-level starts dominating at higher masses of $m_{\gamma_{D}}$.
For $m_{\gamma_{D}} > 1.2$ GeV the main contribution to the scattering events
comes from parton-level channel. In this work the inclusion of parton-level production channel is important as the beam energy considered for 
production of DM is high.
As it is evident from left
(nucleon scattering) and right (electron scattering) panel of figure [\ref{fig3}] that for lower masses of dark photon looking at signatures of
DM scattering with electrons will provide a promising way for the 
observance of DM candidates but if the DP mass is high ($m_{\gamma_{D}} \gtrsim 100$ MeV) then looking at signatures of DM scattering with nucleon
will provide an effective way for the detection of DM candidates.

\begin{figure}[H]
\centering
\includegraphics[width=0.9\linewidth]{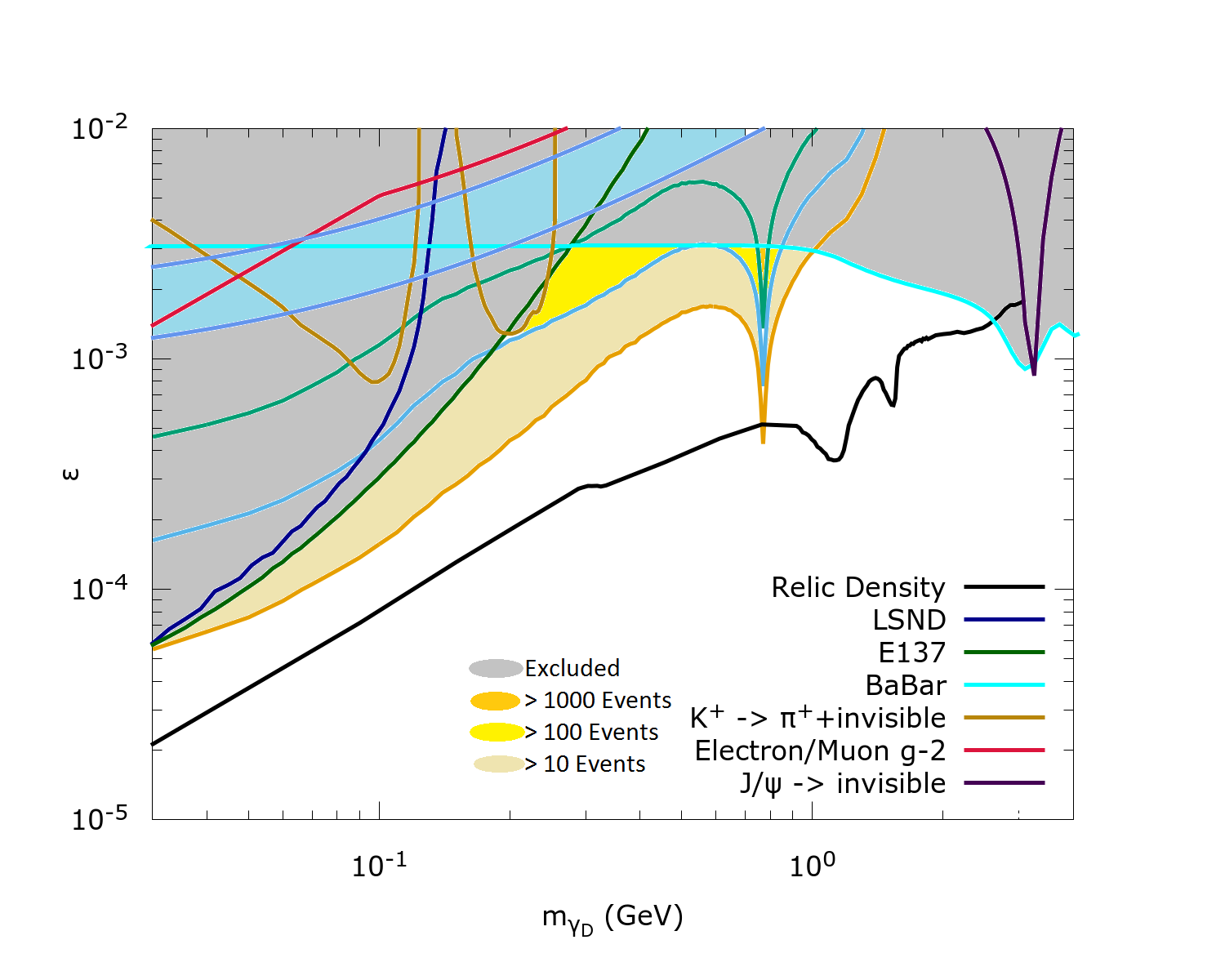}
\caption{The contour sensitivity plot for light dark matter signatures at DUNE in the parameter space of ($\epsilon, m_{\gamma_{D}}$).
These dark matter are produced from distinct channels by using 120 GeV proton beam. Here we have considered $m_{\chi} = 0.01$ GeV,
$\alpha_{D}$ = $0.5$ and POT = $1.1 \times 10^{21}$. In above plot,
the gray regions are excluded by existing constraints, while the yellow
contours indicate 10, 100 and 1000 events
and black line shows thermal relic dark matter (freeze-out).}
\label{ncesens}
\end{figure}

\begin{figure}[H]
\centering
\includegraphics[width=0.9\linewidth]{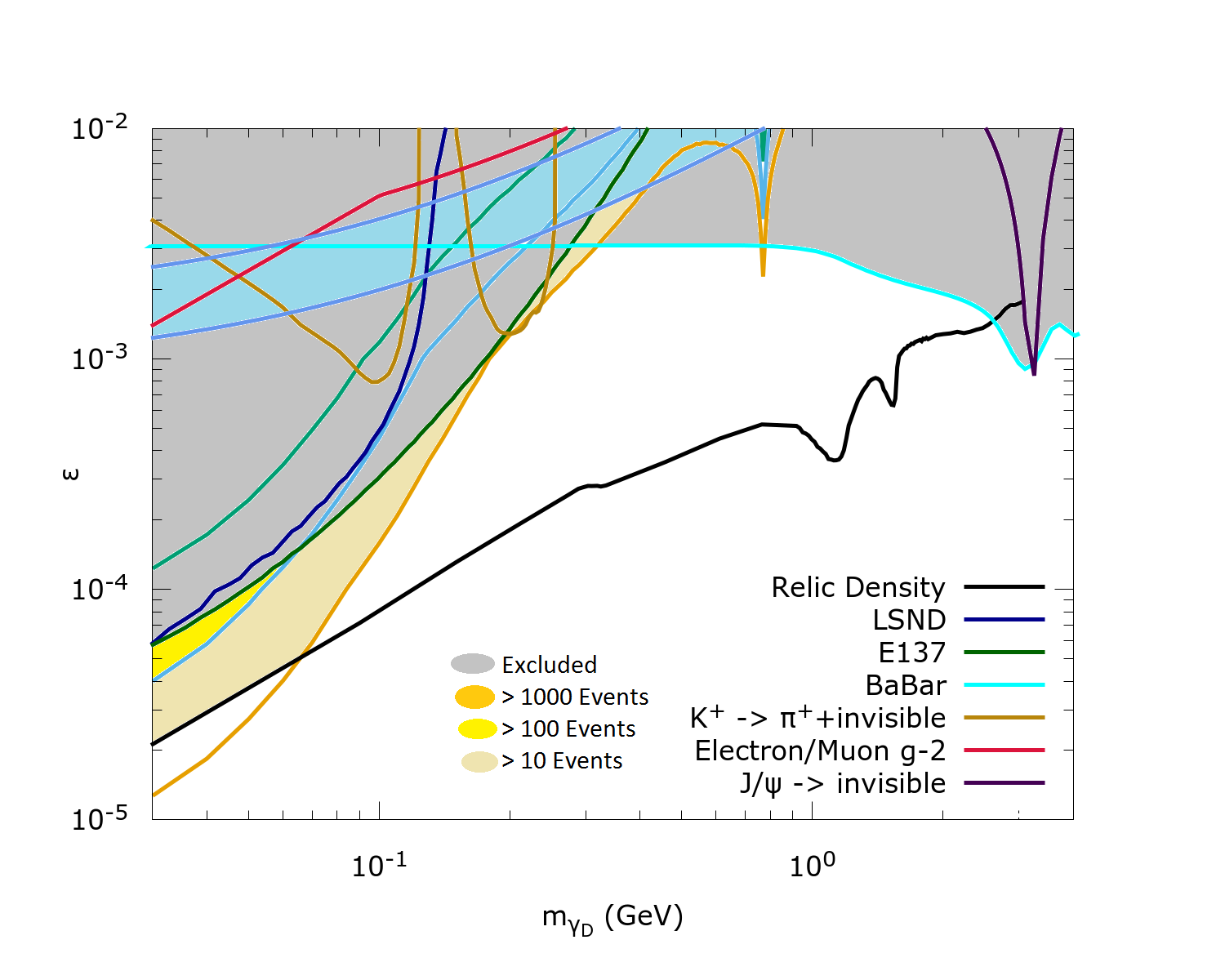}
\caption{The contour sensitivity plot for light dark matter signatures at DUNE in the parameter space of ($\epsilon, m_{\gamma_{D}}$).
These dark matter are produced from distinct channels by using 120 GeV proton beam. Here we have considered $m_{\chi} = 0.01$ GeV,
$\alpha_{D}$ = $0.5$ and POT = $1.1 \times 10^{21}$. In above plot,
the gray regions are excluded by existing constraints, while the yellow
contours indicate 10, 100 and 1000 events
and black line shows thermal relic dark matter (freeze-out).}
\label{esens}
\end{figure}

\begin{figure}[H]
\centering
\includegraphics[width=0.9\linewidth]{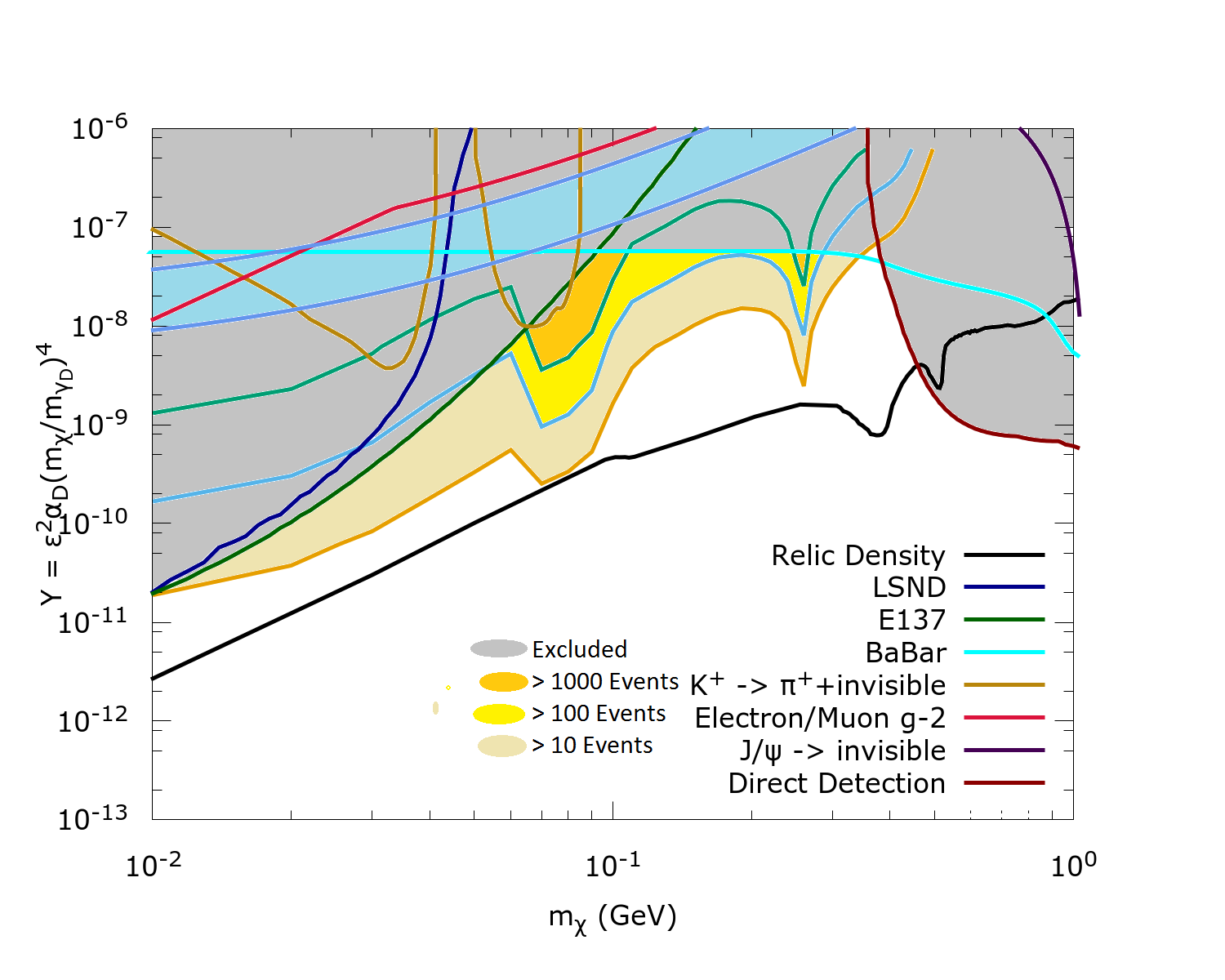}
\caption{The contour sensitivity plot for light dark matter signatures at DUNE in the parameter space of ($Y, m_{\chi}$).
These dark matter are produced from distinct channels by 
 using 120 GeV proton beam. Here we have considered $m_{\gamma_{D}} = 3m_{\chi}$, $\alpha_{D}$ = $0.5$  
  and POT = $1.1 \times 10^{21}$. In above plot,
the gray regions are excluded by existing constraints, while the yellow
contours indicate 10, 100 and 1000 events
and black line shows thermal relic dark matter (freeze-out).}
\label{nceyield}
\end{figure}

\begin{figure}[H]
\centering
\includegraphics[width=0.9\linewidth]{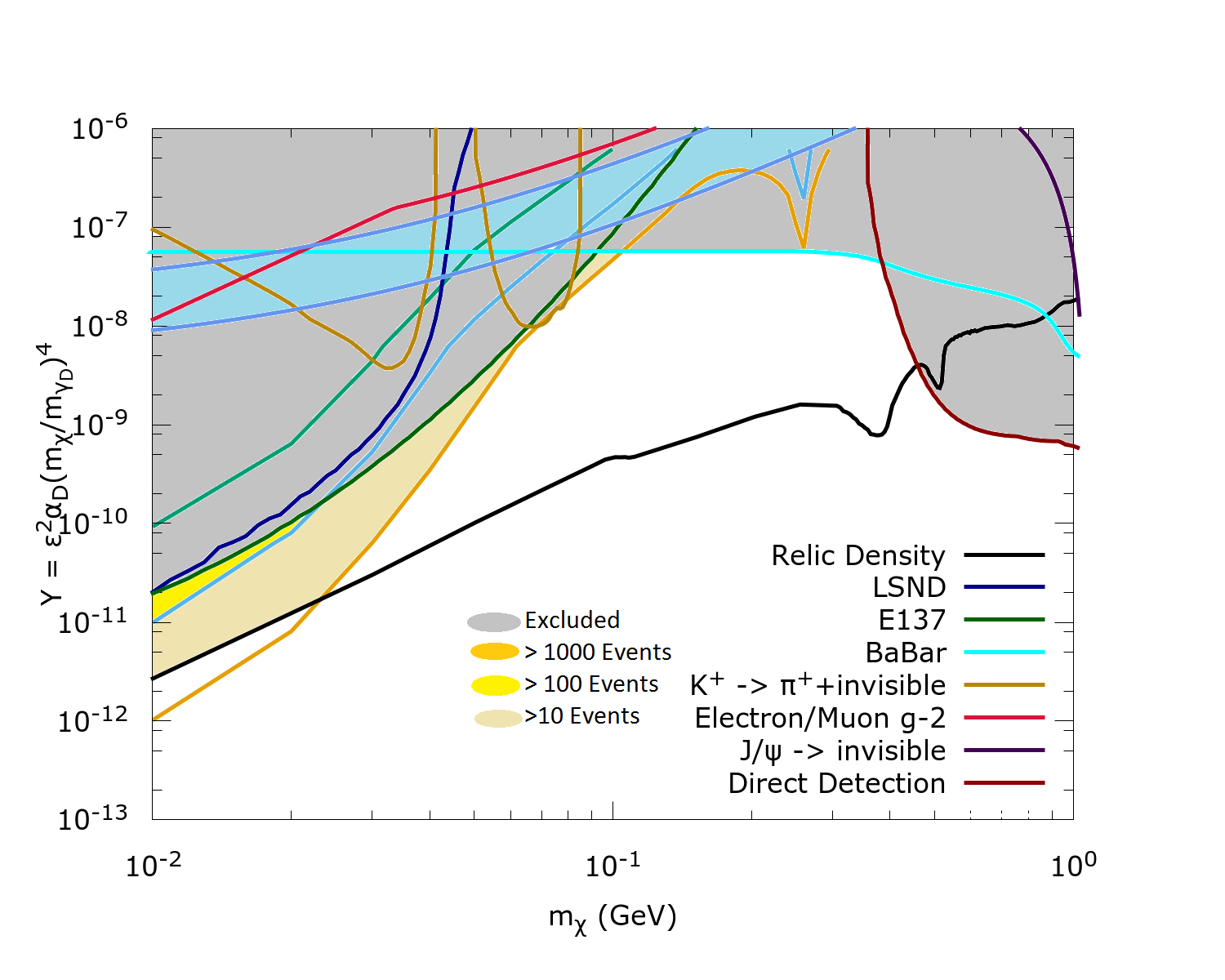}
\caption{The contour sensitivity plot for light dark matter signatures at DUNE in the parameter space of ($Y, m_{\chi}$).
These dark matter are produced from distinct channels by 
 using 120 GeV proton beam. Here we have considered $m_{\gamma_{D}} = 3m_{\chi}$, $\alpha_{D}$ = $0.5$  
  and POT = $1.1 \times 10^{21}$. In above plot,
the gray regions are excluded by existing constraints, while the yellow
contours indicate 10, 100 and 1000 events
and black line shows thermal relic dark matter (freeze-out).}
\label{eyield}
\end{figure}

In figure [\ref{fig3}] the value of kinetic mixing term $\epsilon$ is kept constant at $\epsilon = 10^{-3}$.
To constraints the DM parameters space more tightly the DM parameter $\epsilon$ is allowed to vary between $10^{-2}$ to $10^{-5}$ 
and DP mass is allowed to vary in the range between 0.03 GeV and 2 GeV for the fixed mass of DM. This constraint parameter space $\epsilon$ 
and $m_{\gamma_{D}}$ are shown by
figure [\ref{ncesens}] and [\ref{esens}].
With the help of the thermal relic abundance of DM \cite{c} as shown in figure [\ref{ncesens}] and [\ref{esens}] (black line) realistic 
constraints on parameter
space $\epsilon$ 
and $m_{\gamma_{D}}$ for the fixed mass of DM is explored by DM-nucleon and DM-electron scattering events at DUNE detector. 
For the DM-nucleon scattering (figure [\ref{ncesens}] and [\ref{nceyield}]) we have 
imposed $E_{R} \in [0.1,2]$ GeV cuts on the recoil energy of nucleons. Similarly for DM-electron scattering (figure [\ref{esens}] and [\ref{eyield}])
a stringent forward angle cut 
is imposed on electron i.e. $\theta_{e} \in [0.01,0.02]$ and the energy cut on scattered electron is $E_{e} \in [0.1,2]$ GeV.
The sensitivity contours of DUNE in $\epsilon$ and $m_{\gamma_{D}}$ for elastic scattering of DM with nucleons and electrons
by a considering given mass of DM (10 MeV) are in $\epsilon$
and $m_{\gamma_{D}}$ plane are shown by 10, 100 and 1000 events.
From our analysis we can observe that the DM-electron 
scattering is more sensitive to lower kinetic mixing and lower mass parameter space ($\epsilon, m_{\gamma_{D}}$). Whereas 
DM-nucleon scattering signatures are more sensitivity for kinetic mixing 
$\mathcal{O}(10^{-4} - 10^{-3})$ and higher masses of dark photon $m_{\gamma_{D}}(0.1 - 2 GeV)$ in parameter space ($\epsilon, m_{\gamma_{D}}$).
A sharp increase in DUNE sensitivity is observed at $\sim 800$ MeV mass of $m_{\gamma_{D}}$
which can be attributed to the resonance production at $m_{\gamma_{D}} \sim m_{\rho}(m_{\omega})$ via bremsstrahlung.

 In the figure [\ref{nceyield}] and [\ref{eyield}] DM mass ($m_{\chi}$) varies from 10 MeV to 1000 MeV and kinetic mixing parameter ($\epsilon$) 
 varies from
 $10^{-5}$ to $10^{-1}$. A fixed number of DM scattering events are selected to illustrate the DUNE 
 sensitivity for sub-GeV DM. For the generation of these DM scattering events, $m_{\gamma_{D}} = 3m_{\chi}$
 and fine structure constant of hidden sector $\alpha_{D} = 0.5$ are taken into consideration.
 Since a given mass of dark matter produced by dark photon
 are of different energies hence no linear trend between $m_{\chi}$ and $\epsilon$ are
 observed.
 This parameter space ($Y, m_{\chi}$) is shown in figure [\ref{nceyield}] and [\ref{eyield}] for
 different number of events i.e. 10, 100, 1000 events.
 In this scenario the DM-electron scattering is effective for the study of lower mass of DM ($< 40 MeV$)
 whereas DM-nucleon scattering is effective for larger range and greater values of DM mass.
 A sharp increase in DUNE sensitivity is observed at $\sim 300$ MeV mass of $m_{\chi}$ which can be attributed to the
 resonance production at $m_{\chi} = \frac{m_{\gamma_{D}}}{3} \sim \frac{m_{\rho}(m_{\omega})}{3}$ via bremsstrahlung.

\section{Conclusion:}
Dark matter detection is one of the biggest challenge before the scientific community and seeks input from different branches of science to resolve it.
Cosmological 
and astrophysical evidences of DM needs some new physics to explain the existence of the DM. Thermal relic DM abundance provides conceptually one of the 
simplest and relevant information for the development of DM physics models. Significant
efforts are under way to explore the DM signatures via direct or indirect detection technique. Generally the scattering signatures of DM
mimics the signatures of neutrinos and they adds a significant background to the DM detection. Hence mitigation of neutrino 
background is necessary. There are various ways \cite{7} for the reduction of neutrino background but here we have used beam dump mode technique. 
This beam dump mode
approach is very effective to mitigate the neutrino background, where charged mesons are dumped in the decay volume (which decay into neutrinos).
This approach is recently used by the MiniBooNE experiment \cite{7}.


We have analyzed the simulated study of the sub-GeV DM production and detection at DUNE near detector through direct detection technique at fixed target,
in beam
dump mode. We have tried to capture the sub-GeV DM signatures via DM-nucleon and DM-electron scattering at DUNE near detector.
For this we have explored the DM parameter space ($\epsilon$, $m_{\gamma_{D}}$) and ($Y$, $m_{\chi}$) by taking the thermal relic DM 
density (black line) for scalar DM candidates as benchmark. The fine structure constant of hidden sector is $\alpha_{D} = 0.5$ in our analysis. In 
upcoming work we will try to explore the sensitivity plot for different values of $\alpha_{D}$ too. The results indicates that the 
DM-electron scattering is more efficient at lower masses, whereas DM-nucleon scattering sensitivity
is better in higher masses regime.
The DUNE experiment will be sensitive to the larger parameter space of ($m_{\chi}, \epsilon^{2}\alpha_{D}$) in comparison to other proton beam fixed 
target experiments i.e. LSND and MiniBooNE \cite{61, 102, 103, 104}.
Present constrained on DM parameter space imposed by DUNE is better constrained by prospective constraints on DM parameter space by the different experiments.
We have observed that the sensitivity of DUNE detector for sub-GeV DM is roughly by $50\%$ better as in comparison to 
neutrino experiments.

\subsection{Viability of the model:}
We have focused on the sub-GeV mass range of DM candidates in the proton fixed target experiment to probe the thermal relic DM by considering a model
in which a massive dark photon mediator $\gamma_{D}$, a hypothetical particle belonging to $U(1)_{D}$ gauge group kinetically mixes with the 
ordinary photon of the SM through the kinetic mixing term $\epsilon$.
To address the discrepancy observed between the experimental and theoretical values of the anomalous magnetic moment $(g - 2)$ of leptons,
the model under consideration keeps the window open
(with DUNE)
for handling the $(g - 2)$ anomaly. The mono-photon searches at Belle-II \cite{A} experiment will further decide the viability of the vector portal.
The collider searches for missing energy and momentum can further add information to the dark matter model.
The forthcoming e-ASTROGAM \cite{B} a gamma ray telescope will offer important platform to discover dark matter particles with masses below $\sim 10$ MeV
and check the viability of the model. Further CRESST-II experiment pushes down the mass of DM below 1 GeV in the direct detection technique.
In this way we can see that the fusion of all approaches will allow us to pin down the dark matter parameter space.

\begin{acknowledgement}
We would like to thank Prof. Raj Gandhi for his valuable help and discussions.
\end{acknowledgement}\vspace{-10mm}


\begin{thebibliography}{99.}%
 
\bibitem{72}
  Timothy Cohen et al., Wino Dark Matter Under Siege, (2013), arXiv:1307.4082v3 [hep-ph].
  
\bibitem{73}
  JiJi Fan and Matthew Reece, In Wino Veritas? Indirect Searches Shed Light on Neutralino Dark Matter, (2013), arXiv:1307.4400v1 [hep-ph].

\bibitem{74}
  A. Hryczuk et al., Indirect Detection Analysis: Wino Dark Matter Case Study, (2014), arXiv:1401.6212v2 [astro-ph.HE].
  
\bibitem{71}
  G. Angloher et al., Results from 730 kg days of the CRESST-II Dark Matter Search, (2012), arXiv:1109.0702v1 [astro-ph.CO].
  
\bibitem{75}
  B. Bhattacherjee et al., Phenomenology of Light Fermionic Asymmetric Dark Matter, (2013), arXiv:1306.5878v2 [hep-ph].
  
\bibitem{76}
  M. Blennow et al., Asymmetric Dark Matter and Dark Radiation, (2012), arXiv:1203.5803v2 [hep-ph].
  
\bibitem{77}
  D. E. Kaplan et al., Asymmetric Dark Matter, (2009), arXiv:0901.4117v1 [hep-ph].
  
\bibitem{78}
  Ian-Woo Kim and K. M. Zurek, Flavor and Collider Signatures of Asymmetric Dark Matter, (2013), arXiv:1310.2617v1 [hep-ph].
  
\bibitem{79} 
 E. Aprile et al., (XENON100 Collaboration), Dark Matter Results from 225 Live Days of XENON100 Data, (2012), arXiv:1207.5988v2 [astro-ph.CO].
 
\bibitem{80}
G. Angloher et al., Results on light dark matter particles with a low-threshold CRESST-II detector (CRESST Collaboration), 
Eur. Phys. J. C76 no. 1, (2016) 25, arXiv:1509.01515 [astro-ph.CO].

\bibitem{81}
R. Agnese et al., (SuperCDMS Collaboration), WIMP-Search Results from the Second CDMSlite Run, (2016), arXiv:1509.02448v2 [astro-ph.CO].

\bibitem{82}
A. Tan et al., (PandaX-II Collaboration), Dark Matter Results from First 98.7 Days of Data from the PandaX-II Experiment, (2016), arXiv:1607.07400v3
[hep-ex].

\bibitem{83}
D. S. Akerib et al., Results from a search for dark matter in the complete LUX exposure, (2017), arXiv:1608.07648v3 [astro-ph.CO].

\bibitem{84}
J. Alexander et al., Dark Sectors 2016 Workshop: Community Report, (2016), arXiv:1608.08632v1 [hep-ph].

\bibitem {4}
B.W. Lee and S. Weinberg, Cosmological Lower Bound on Heavy-Neutrino Masses, Phys. Rev. Lett. 39 (1977) 165.  
  
\bibitem{7}
R. Dharmapalan et al., (The MiniBooNE Collaboration), Low Mass WIMP Searches with a Neutrino Experiment:A Proposal for Further MiniBooNE Running, (2012), 
arXiv:1211.2258v1 [hep-ex].
              
\bibitem{1} 
R. Acciarri et al., (The DUNE Collaboration), Long-Baseline Neutrino Facility (LBNF) and Deep Underground Neutrino Experiment (DUNE), 
(2016), arXiv:1601.02984v1 [physics.ins-det]. 

\bibitem{61}
Patrick deNiverville et al., Light dark matter in neutrino beams: production modelling and 
scattering signatures at MiniBooNE, T2K and SHiP, arXiv:1609.01770v3 [hep-ph].

\bibitem{86}
Anirban Das and Basudeb Dasgupta, Selection Rule for Enhanced Dark Matter Annihilation, (2017), arXiv:1611.04606v3 [hep-ph]
 
\bibitem{43}
N. Prantzos et al., The 511 keV emission from positron annihilation in the Galaxy, (2010), arXiv:1009.4620v1 [astro-ph.HE].

\bibitem{44}
L. Bouchet et al., On the morphology of the electron-positron annihilation emission as seen by SPI/INTEGRAL, (2010), arXiv:1007.4753v1 [astro-ph.HE].

\bibitem{50}
D. Hooper and K. M. Zurek, Natural supersymmetric model with MeV dark matter, Phys.Rev.D77, 087302 (2008).

\bibitem{51}
 J. L. Feng and J. Kumar, The WIMPless Miracle: Dark Matter Particles without Weak-scale Masses or Weak Interactions, (2008),
 arXiv:0803.4196v3 [hep-ph].
 
\bibitem{52}
 J.L. Feng, H. Tu, and H.B. Yu, Thermal Relics in Hidden Sectors, (2008), arXiv:0808.2318v3 [hep-ph].

 \bibitem{38}
P. Ade et al., (Planck Collaboration), Planck 2013 results. I. Overview of products and scientific results , (2013), arXiv:1303.5062v2 [astro-ph.CO].

\bibitem{17}
 D. B. Kaplan, Single Explanation for Both Baryon and Dark Matter Densities, Phys. Rev. Lett.68, 741 (1992).
 
\bibitem{18}
  G. R. Farrar and G. Zaharijas, Dark Matter and the Baryon Asymmetry of the Universe, (2004), arXiv:hep-ph/0406281v3.
    
\bibitem{36}
 K. M. Zurek, Asymmetric Dark Matter: Theories, Signatures, and Constraints, (2013), arXiv:1308.0338v2 [hep-ph].
  
\bibitem{37}
T. Cohen, D.J. Phalen, A. Pierce and K. M. Zurek, Asymmetric Dark Matter from a GeV Hidden Sector, (2010), arXiv:1005.1655v1 [hep-ph].

\bibitem{15}
B. Holdom, Two U(1)'s and Epsilon Charge Shifts, Phys. Lett. B166 (1986)196.

\bibitem{16}
L. B. Okun, Limits on electrodynamics: paraphotons?, (1982), Zh. Eksp. Teor. Fiz. 83,892-898.

\bibitem{11} 
M. Pospelov, A. Ritz and M. B. Voloshin, Secluded WIMP Dark Matter, (2007), arXiv:0711.4866v1 [hep-ph].

\bibitem{k}
J. P. Lees et al. (The BABAR Collaboration), Search for Invisible Decays of a Dark Photon Produced in $e^{+}e^{-}$ Collisions at BABAR, (2017)
arXiv:1702.03327v2  [hep-ex].

\bibitem{l}
D. Banerjee et al. (NA64), Search for vector mediator of Dark Matter production in invisible decay mode, (2017), arXiv:1710.00971v2 [hep-ex].

\bibitem{64}
 E. Izaguirre, G. Krnjaic, P. Schuster and N. Toro, Accelerating the Discovery of Light Dark Matter, (2015), arXiv:1505.00011v1 [hep-ph].

\bibitem{88}
B. Batell, M. Pospelov and A. Ritz, Exploring Portals to a Hidden Sector Through Fixed Targets, (2009), arXiv:0906.5614v2 [hep-ph].

\bibitem{54}
M. Bonesini et al., On particle production for high energy neutrino beams, (2001), arXiv:hep-ph/0101163v3.

\bibitem{a}
S. Teis et al., Pion-Production in Heavy-Ion Collisions at SIS energies, (1996), arXiv:nucl-th/9609009v1.

\bibitem{67}
  M. Schwartz, Quantum Field Theory and the Standard Model, (Cambridge University Press), (2014).

\bibitem{68}
  Y. Kahn, G. Krnjaic, J. Thaler and M. Toups, DAE$\delta$ALUS and Dark Matter Detection, (2014), arXiv:1411.1055v3 [hep-ph].  
  
\bibitem{b}
Kevin J. Kelly and Yu-Dai Tsai, Proton Fixed-Target Scintillation Experiment to Search for Minicharged Particles, (2018), arXiv:1812.03998v2 [hep-ph].
  
\bibitem{56}
Johannes Bl$\ddot{u}$mlein and J$\ddot{u}$rgen Brunner, New Exclusion Limits on Dark Gauge Forces from Proton Bremsstrahlung in Beam-Dump Data, 
(2013), arXiv:1311.3870v1 [hep-ph].

\bibitem{57}
D. Gorbunov, A. Makarov, and I. Timiryasov, Decaying light particles in the SHiP experiment:Signal rate estimates for hidden photons,
(2015), arXiv:1411.4007v2 [hep-ph]. 
 
\bibitem{58}
A. Faessler, M. I. Krivoruchenko, and B. V. Martemyanov, Once more on electromagnetic form factors of nucleons in extended vector meson dominance model,
(2009), arXiv:0910.5589v1 [hep-ph].

\bibitem{59}
P. deNiverville, D. McKeen, and A. Ritz, Signatures of sub-GeV dark matter beams at neutrino experiments, (2012), arXiv:1205.3499v1 [hep-ph].
 
\bibitem{60}
P. M. Nadolsky et al., Implications of CTEQ global analysis for collider observables, (2008), arXiv:0802.0007v3 [hep-ph].

\bibitem{c}
P. deNiverville, M. Pospelov, and A. Ritz, Observing a light dark matter beam with neutrino experiments, (2013), arXiv:1107.4580v3 [hep-ph].

\bibitem{62}
 L.A. Ahrens et al., Measurement of neutrino-proton and antineutrino-proton elastic scattering, Phys. Rev. D35, 785(1987).

\bibitem{d}
B. Batell, P. deNiverville, D. McKeen, M. Pospelov, and A. Ritz, Leptophobic Dark Matter at Neutrino Factories, (2014), arXiv:1405.7049v1 [hep-ph].

\bibitem{63}
 High-Pressure Argon gas TPC Option for the DUNE Near Detector, DUNE HPgTPC WG, Fermi National Accelerator Laboratory, Box 500, 
 Batavia, IL 60510-5011, USA.
 
\bibitem{66}
 Garrett BROWN, Sensitivity Study for Low Mass Dark Matter Search at DUNE, (2018).
 
\bibitem{87}
Gert H$\ddot{u}$tsi et al., WMAP7 and future CMB constraints on annihilating dark matter:implications for GeV-scale WIMPs, (2011),
arXiv:1103.2766v3 [astro-ph.CO].
 
\bibitem{97}
Brian Henning and Hitoshi Murayama, Constraints on Light Dark Matter from Big Bang Nucleosynthesis, (2012), arXiv:1205.6479v1 [hep-ph].
 
\bibitem{88a}
M. Markevitch et al., Direct constraints on the dark matter self-interaction cross-section from the merging galaxy cluster 1E0657-56,
(2004), arXiv:astro-ph/0309303v2.

\bibitem{89}
J. Miralda-Escude, A Test of the Collisional Dark Matter Hypothesis from Cluster Lensing, (2001), arXiv:astro-ph/0002050v2.

\bibitem{91}
P. Fayet, Invisible $\Upsilon$ decays into Light Dark Matter, (2010), arXiv:0910.2587v2 [hep-ph].

\bibitem{92}
R.M. Bionta et al., Observation of a Neutrino Burst in Coincidence with Supernova 1987A in the Large Magellanic Cloud,
Phys.Rev.Lett. 58, 1494 (1987); \\
K. Hirata et al., Observation of a Neutrino Burst from the Supernova SN1987A,
(Kamiokande-II), Phys.Rev.Lett. 58, 1490 (1987); \\
E. Izaguirre, G. Krnjaic, P. Schuster, and N. Toro, Testing GeV-Scale
Dark Matter with Fixed-Target Missing Momentum Experiments, (2015), arXiv:1411.1404v3 [hep-ph].

\bibitem{93}
G. Angloher et al., (CRESST Collaboration), Results on light dark matter particles with a low-threshold CRESST-IIdetector, Eur. Phys. J. C76, 25 (2016), arXiv:1509.01515 [astro-ph.CO].

\bibitem{94}
R. Agnese et al., (SuperCDMS Collaboration), WIMP-Search Results from the Second CDMSlite Run, (2016), arXiv:1509.02448v2 [astro-ph.CO].

\bibitem{95}
B. A. Dobrescu and C. Frugiuele, Hidden GeV-scale interactions of quarks, (2014), arXiv:1404.3947v2 [hep-ph].

\bibitem{96}
B. Batell, R. Essig, and Z. Surujon, Strong Constraints on Sub-GeV Dark Sectors from SLAC Beam Dump E137, Phys.Rev.Lett. 113, 171802 (2014), arXiv:1406.2698 [hep-ph].

\bibitem{m}
M. Pospelov, Secluded U(1) below the weak scale, (2008), arXiv:0811.1030v1 [hep-ph].

\bibitem{n}
A. V. Artamonov et al. (E949 Collaboration), Study of the decay $K^{+} \to \pi\nu\bar{\nu}$ in the momentum region $140< P_{\pi} <199$ MeV/c,
(2009), arXiv:0903.0030v1 [hep-ex].

\bibitem{o}
R. Bouchendira et al., New determination of the fine structure constant and test of the quantum electrodynamics, (2010),
arXiv:1012.3627v1 [physics.atom-ph].

\bibitem{p}
T. Aoyama et al., Tenth-Order QED Contribution to the Electron $g−2$ and an Improved Value of the Fine Structure Constant, (2012),
arXiv:1205.5368v2 [hep-ph].

\bibitem{q}
D. Hanneke, S. F. Hoogerheide, and G. Gabrielse, Cavity Control of a Single-Electron Quantum Cyclotron: Measuring the Electron Magnetic Moment,
(2010), arXiv:1009.4831v1 [physics.atom-ph].

\bibitem{100}
A.A. Aguilar-Arevalo et al., (The MiniBooNE-DM Collaboration), Dark matter search in nucleon, pion, and electron channels from a proton beam dump
with MiniBooNE, (2018), arXiv:1807.06137v2 [hep-ph].

\bibitem{101}
A. A. Aguilar-Arevalo et al., (The MiniBooNE-DM Collaboration), A Combined $\nu_{\mu} \to \nu_{e}$ \& $\bar{\nu}_{\mu} \to \bar{\nu}_{e}$
Oscillation Analysis of the MiniBooNE Excesses, (2012), arXiv:1207.4809v2 [hep-ex].

\bibitem{102}
Claudia Frugiuele, Probing sub-GeV dark sectors via high energy proton beams at LBNF/DUNE and MiniBooNE, Phys. Rev., D 96, 015029 (2017).

\bibitem{103}
Valentina De Romeri, Kevin J. Kelly and Pedro A. N. Machado, DUNE-PRISM sensitivity to light dark matter, Phys. Rev., D 100, 095010 (2019).

\bibitem{104}
Pilar Coloma, Bogdan A. Dobrescu, Claudia Frugiuele and Roni Harnik, Dark matter beams at LBNF, JHEP 04 (2016) 047.

\bibitem{105}
S. Profumo et al., An Introduction to Particle Dark Matter, (2019), arXiv:1910.05610v1 [hep-ph].

\bibitem{A}
R. Essig et al., Constraining Light Dark Matter with Low-Energy $e^{+}e^{-}$ Collider , (2015), arXiv:1309.5084v2 [hep-ph].

\bibitem{B}
Ma$\acute{i}$ra Dutra et al., MeV Dark Matter Complementarity and the Dark Photon Portal, (2018), arXiv:1801.05447v2 [hep-ph].

\end{thebibliography}
\end{document}